\documentclass[journal,draftclsnofoot,onecolumn,12pt]{IEEEtran}
\usepackage{wrapfig,booktabs,fancyhdr,amsmath,amsfonts,tabularx,numprint}
\usepackage{cite,bm,bbm,amssymb,amsthm,url,multirow,times,enumitem,comment}
\usepackage{mathtools,siunitx,balance,tikz,adjustbox,graphicx,array}
\usepackage[font=footnotesize]{subcaption}
\usepackage[linesnumbered,ruled]{algorithm2e}
\usepackage[colorlinks=true, linkcolor=black, citecolor=blue, urlcolor=blue]{hyperref}

\newcommand{\by}{\times}

\newcommand{\ba}{\begin{array}}
\newcommand{\ea}{\end{array}}
\newcommand{\sinc}{\text{sinc}}

\newcommand{\E}[1]{\mathbb{E}\left[ #1 \right]}

\DeclareMathAlphabet{\mathpzc}{OT1}{pzc}{m}{it}
\DeclareMathOperator*{\argmin}{\arg\!\min}
\DeclareMathOperator*{\argmax}{\arg\!\max}
\usepackage{color}

\begin{document}

\def\SNR{\mathrm{SNR}}
\def\SINR{\mathrm{SINR}}
\def\SIR{\mathrm{SIR}}
\def\MMSE   {\mbox{\scriptsize \sf MMSE}}
\def\sir {\mbox{\scriptsize\sf SIR}}
\def\snr {\mbox{\scriptsize\sf SNR}}
\def\sinr {\mbox{\scriptsize\sf SINR}}
\newcommand{\SIC}{\mathrm {SIC}}

\newcommand{\pout}{{P_{\rm out}}}

\def\sinc{\mbox{\textrm sinc}}
\def\dd{\mathrm{d}}
\newcommand{\ceil}[1]{\lceil #1\rceil}
\def\argmin{\operatorname{arg~min}}
\def\argmax{\operatorname{arg~max}}

\newcommand{\PP}{\mathbb{P}}
\def\E{\mathbb{E}}
\newcommand{\Var}{{\rm Var}}

\def\Es{\mathcal{E}_s}
\def\Eb{\mathcal{E}_b}

\def\bydef{:=}
\def\ba{{\mathbf{a}}}
\def\bb{{\mathbf{b}}}
\def\bc{{\mathbf{c}}}
\def\bd{{\mathbf{d}}}
\def\bee{{\mathbf{e}}}
\def\bff{{\mathbf{f}}}
\def\bg{{\mathbf{g}}}
\def\bh{{\mathbf{h}}}
\def\bi{{\mathbf{i}}}
\def\bj{{\mathbf{j}}}
\def\bk{{\mathbf{k}}}
\def\bl{{\mathbf{l}}}
\def\bm{{\mathbf{m}}}
\def\bn{{\mathbf{n}}}
\def\bo{{\mathbf{o}}}
\def\bp{{\mathbf{p}}}
\def\bq{{\mathbf{q}}}
\def\br{{\mathbf{r}}}
\def\bs{{\mathbf{s}}}
\def\bt{{\mathbf{t}}}
\def\bu{{\mathbf{u}}}
\def\bv{{\mathbf{v}}}
\def\bw{{\mathbf{w}}}
\def\bx{{\mathbf{x}}}
\def\by{{\mathbf{y}}}
\def\bz{{\mathbf{z}}}
\def\b0{{\mathbf{0}}}

\def\bA{{\mathbf{A}}}
\def\bB{{\mathbf{B}}}
\def\bC{{\mathbf{C}}}
\def\bD{{\mathbf{D}}}
\def\bE{{\mathbf{E}}}
\def\bF{{\mathbf{F}}}
\def\bG{{\mathbf{G}}}
\def\bH{{\mathbf{H}}}
\def\bI{{\mathbf{I}}}
\def\bJ{{\mathbf{J}}}
\def\bK{{\mathbf{K}}}
\def\bL{{\mathbf{L}}}
\def\bM{{\mathbf{M}}}
\def\bN{{\mathbf{N}}}
\def\bO{{\mathbf{O}}}
\def\bP{{\mathbf{P}}}
\def\bQ{{\mathbf{Q}}}
\def\bR{{\mathbf{R}}}
\def\bS{{\mathbf{S}}}
\def\bT{{\mathbf{T}}}
\def\bU{{\mathbf{U}}}
\def\bV{{\mathbf{V}}}
\def\bW{{\mathbf{W}}}
\def\bX{{\mathbf{X}}}
\def\bY{{\mathbf{Y}}}
\def\bZ{{\mathbf{Z}}}

\title{6G Takes Shape}
\author{Jeffrey G. Andrews, Todd E. Humphreys, and Tingfang Ji
\thanks{Jeffrey G. Andrews (jandrews@ece.utexas.edu) and Todd E. Humphreys (todd.humphreys@utexas.edu) are with 6G@UT in the Wireless Networking and Communications Group (WNCG) at The University of Texas at Austin, Austin, TX, USA. Tingfang Ji is with Qualcomm Technologies, Inc., San Diego, CA.}
\thanks{The opinions and perspectives expressed in this paper are those of the authors, not those of Qualcomm or the University of Texas at Austin.}
\thanks{Please cite as appearing in \emph{IEEE BITS the Information Theory Magazine}, Special Issue on 6G, Dec. 2024.}}

\maketitle

\begin{abstract}
  The contours of 6G---its key technical components and driving
  requirements---are finally coming into focus. Through twenty questions and
  answers, this article defines the important aspects of 6G across four
  categories.  First, we identify the key themes and forces driving the
  development of 6G, and what will make 6G unique.  We argue that 6G
  requirements and system design will be driven by (i) the tenacious pursuit of
  spectral (bits/Hz/area), energy (bits/Joule), and cost (bits/dollar)
  efficiencies, and (ii) three new service enhancements:
  sensing/localization/awareness, compute, and global broadband/emergency
  connectivity.  Second, we overview the important role of spectrum in 6G, what
  new spectrum to expect in 6G, and outline how the different bands will be used
  to provide 6G services.  Third, we focus our attention on the 6G physical
  layer, including waveforms, MIMO advancements, and the potential use of deep
  learning.  Finally, we explore how global connectivity will be achieved in 6G,
  through non-terrestrial networks as well as low-cost network expansion via
  disaggregation and O-RAN.  Although 6G standardization activities will not
  begin until late 2025, meaning this article is by definition speculative, our
  predictions are informed by several years of intensive research and
  discussions.  Our goal is to provide a grounded perspective that will be
  helpful to both researchers and engineers as we move into the 6G era.
\end{abstract}

\section{Introduction: What this paper is about}
The goal of this article is to provide a realistic view of what the 6G standard
and cellular network is likely to look like by the early 2030s.  A great deal of
intriguing research has been done on 6G technologies, nearly from the moment 5G
standards were released.  This paper does not attempt to summarize all such
work, nor predict or discuss every potential aspect of 6G.  Instead, it
summarizes what we recognize as an emerging but mostly unacknowledged consensus
on many aspects of 6G. On other aspects, there may not yet be consensus, but
based on our own research, our understanding of others' research, and our
perspective on the industry and its needs, we nonetheless believe some outcomes
are all but inevitable.

The goal of this paper is to share these developments and predictions, and the
rationales behind them.  We structure the paper around 20 technical and
business-focused questions on 6G subdivided into four categories.  We briefly
summarize these questions here, then package our key theses and predictions
within the answers to the questions.

\textbf{The Case for 6G.} First, we consider the key rationales and forces
driving development of 6G.  On one level, there is something inevitable about a
new G being introduced every decade: the entire cellular industry has been built
around this assumption since at least the 2G era.  But this assumption is under
increasing scrutiny.  At the very, least, we believe the meaning of 6G is likely
to be quite different from that of previous Gs.  We explore the big picture of
6G through the following questions:

\begin{enumerate}
\item There is a widespread sense that 5G has been a disappointment.  Why is
  that?  Is it true?
\item Based on lessons learned from 5G, what should 6G do differently?
\item Besides ``fixing 5G,'' what new use cases, services, and applications will drive 6G?
\item What will be the new value proposition for 6G?
\item Will 6G will be backward-compatible with 5G?
\item Will 6G be the last G?
\end{enumerate}

\textbf{Spectrum in 6G.}  Already, 5G more than doubled the total amount of
spectrum in use for cellular communication.  We expect a similar doubling for
6G, with cellular communication increasingly expanding out of its heartland
spectrum (carrier frequencies under 2 GHz) into a diverse mix of different bands
with different characteristics.  Some of these bands will require at least
modest forms of spectrum sharing, or other new fundamental technology
advancements.  This raises several interesting questions:

\begin{enumerate}[resume]
\item Which spectrum bands will have a major impact on 6G?
\item Will millimeter wave actually be used in 6G?
\item How will the different spectral bands in 6G be utilized? 
\item What roles will Terahertz (THz) spectrum and Reflective Intelligent
  Surfaces (RIS) play?
\item How will spectrum be allocated in 6G?  Will dynamic spectrum sharing
  across operators and systems become the norm?
\end{enumerate}

\textbf{The 6G PHY.}  Third, we discuss the 6G physical layer (PHY), including
multiple access.  Historically, this has been the defining aspect of each new G.
The physical layer has also yielded the most novel aspects for each G, and has
broken backward compatibility, thus requiring new radios and hardware.  We
structure this section around the following questions spanning fundamental
aspects of the PHY and the use of ML in 6G:
\begin{enumerate}[resume]
\item Will there be a new waveform for 6G beyond OFDM/OFDMA?
\item Will there be new types of coding, modulation, or duplexing? 
\item Will we see any new types of MIMO in 6G?
\item Will the 6G PHY be replaced by neural networks and run on GPUs?
\item What are the key roles for machine learning in 6G?
\end{enumerate}

\textbf{True Global Connectivity.} Fittingly, the last aspect we cover is the
``final frontier'': the long-held dream of providing true broadband connectivity
to everyday cellular devices anywhere on Earth.  This includes tighter
integration with the emerging dense low-Earth orbit (LEO) constellations,
including possible direct-to-handset communication (D2C) at least as a backup to
terrestrial networks.  It also implies considerable network densification in the
developing world, which may be enabled by lower-cost deployments.  The final
four questions are
\begin{enumerate}[resume]
\item Can we really achieve global broadband coverage directly to smartphones
   with emerging LEO satellite constellations, or is that a pipe dream?
\item What role will the 6G cellular standard play in D2C?
\item What about other non-terrestrial network (NTN) paradigms such as High
  Altitude Platform Stations (HAPS) for achieving global coverage?
\item Is O-RAN going to be transformative, or will the operators stick with a
  more vertically integrated system?
\end{enumerate}

We offer some strong and perhaps unpopular opinions in this paper.  Some are
optimistic, some are pessimistic.  We are sure many readers will find themselves
in fierce disagreement with some of our predictions and arguments.  No doubt
some of our predictions will prove incorrect in the end.  Nevertheless, we offer
this unflinching perspective to the literature on 6G to encourage thoughtful
debate.  We share the excitement of 6G finally starting to take shape.

\section{The case for 6G}

\newcounter{Qnum}
\stepcounter{Qnum}
An engineer cannot design a good system if the goals and requirements for that
system are unclear. This includes understanding the limitations of the
existing technology.  Thus, we start with 5G.

\subsection{Question \theQnum: There is a widespread sense that 5G has been a
  disappointment.  Why is that?  Is it true?}
\refstepcounter{Qnum} It is at least half true.  

First, the positive: 5G has been largely successful in terms of many objective
metrics, and is still in its adolescence. 5G smartphones are only now beginning
to become a dominant fraction of the market, with over 2 billion shipped: an
uptake rate faster than LTE, nevertheless.  Roughly half the operators globally
are still in 5G Non Stand Alone (NSA) mode, meaning they rely on an LTE carrier
and use 4G core networks.  Perhaps surprisingly, some of the more advanced 5G
deployments are in developing countries such as India, which in many cases
leapfrogged LTE straight to 5G.  New IoT use cases are being enabled on a daily
basis, utilizing Reduced Capability (RedCap) and enhanced RedCap (eRedCap)
modems, also known as 5G NR-light.  RedCap was not even introduced until Release
17 -- the third 5G release -- in 2022.  In short, 5G is still growing and
evolving rapidly, in terms of deployments and use cases.  Another success story
in 5G has been the rapid growth of fixed wireless access (FWA) services, which
are now used by tens of millions of people as their primary home Internet
connection. Further, 5G was designed to be flexible and accommodate a wide
variety of use cases and deployments, providing a solid platform for future
growth.

\textbf{Slow rollout of new verticals.} That said, the sense of disappointment
around 5G is real, and extends from many corners of the wireless industry to the
everyday consumer.  There were bold expectations that 5G would provide a
revolutionary leap forward in the quantity and quality of cellular connectivity,
and quickly unleash many new IoT verticals and low latency use cases
\cite{Fet14}.  This has not happened so far.  One glaring example is the hype
around connected vehicles and driverless cars in the mid-to-late 2010s
\cite{lu2014connectedVehSaC,uhlemann2016connected}, which has not come to pass.  Other examples included grand visions of 5G-enabled “smart factories” and “private networks.”  These are in fact happening to some extent, but were always going to be niche use cases and not capable of driving economies of scale.

\textbf{Millimeter Wave has disappointed in 5G.}  There has also been some
reckonings on the technical side.  The most discussed case is millimeter wave
(mmWave) communications, widely touted as the path to Gbps connectivity in 5G
\cite{Rap2013ItWillWork}.  Yet it was always unrealistic to expect a big early
impact from mmWave.  First, due to severe penetration losses, it is primarily an
outdoor-to-outdoor technology, whereas roughly 80\% of cellular data consumption
today takes place indoors.  Second, mmWave requires significant UE-side support
via multiple antenna arrays with integrated mmWave RF front-ends, which are
costly and not yet widely available.  Third, mmWave requires dense network
deployments, which are expensive.  Meanwhile, the new C-band deployments for 5G
have been a clear success: they now carry a large fraction of 5G traffic and
offer considerable spare capacity in most markets, further undercutting the
rationale for aggressive mmWave deployments .  Nonetheless, we believe mmWave's
prominence will rise in 6G for both technical and business reasons. We discuss
mmWave's future in Sec. \ref{sec:mmWave}.

\textbf{Overly quick and reliable is not a virtue}.  A second 5G disappointment
has been the scant use of the URLLC modes, in part due to technical
misunderstandings about the costs of providing URLLC.  The 5G network remains a
smartphone-oriented network carrying primarily best-effort data traffic, such as
stored video.  The most aggressive URLLC modes, such as downlink pre-emption
(puncturing), are not spectrally or power efficient, and the rationale for such
an approach is dubious, especially in the uplink.  Instead, with shorter slot
times (due in part to wider OFDM subcarriers), increased bandwidth, and faster
feedback cycles, the baseline latency in 5G is already quite good. Consider: the
LTE slot duration is 1 msec whereas the 5G slot durations are 0.5 msec and 0.125
msec for C-band and mmWave, respectively, yielding single-digit millisecond
over-the-air (OTA) latency in 5G. Moreover, the latency of a 5G connection is
often limited by the \emph{transport} latency in the core network, which is
mostly a function of delays between the server and the RAN (i.e., the BS).  This
transport latency can be in the tens of msec.  Thus, further reduction of the
OTA latency leads to rapidly diminishing returns, at high cost.

\textbf{Power consumption and cost.}  5G operators are struggling with
unprecedented power consumption: electricity is a significant percentage (e.g.,
a quarter) of their operating expenses.  Along with the cost of 5G spectrum,
they are facing a mountain of debt, which has suppressed investments in network
expansion.  Emerging environmental regulations such as carbon caps or taxes will
further penalize high network power consumption in the future.  Operators will
be reluctant to deploy 6G unless it can help them address these issues.

\subsection{Question \theQnum: Based on lessons learned from 5G, what should 6G
  do differently?}
\refstepcounter{Qnum} Based on experience with 5G, we expect 6G to adopt a laser
focus on energy efficiency and cost reduction, which will demand novel technical
solutions.  These two new pillars of 6G will require a significant departure
from the pre-6G design mindset, which was focused on delivering ever-better
communications KPIs: better spectral efficiency, lower latency, higher
reliability, and higher peak and median throughput.  5G grouped these KPIs into
three distinct classes of use cases---distilled into the vertices of the ``5G
triangle''-- which are: (i) enhanced Mobile Broadband (eMBB), (ii) massive
Machine Type Communications (mMTC), and (iii) Mission Critical, or URLLC.

\textbf{6G's communication KPIs will not differ much from 5G's}.  We do not
foresee 6G pursuing significantly stronger requirements in terms of
communications KPIs, unlike some have predicted
\cite{giordani2020toward,jiang2021roadComp,wang2023roadOn} and unlike what is
implied by the popular ITU ``6G Hexagon.''  We view the communication aspects of
the 6G Hexagon -- which further expands upon 5G's eMBB, URLLC and mMTC -- as a
slightly mismatched descriptor of 6G and what it will accomplish.  Instead, we
see the targeted 5G communication KPIs as sufficiently aggressive---maybe even
overly aggressive---for nearly all the envisioned 6G-era use cases.  Consider
5G's attempt at URLLC services offering both low roundtrip latency (approaching
1 msec) and high reliability (e.g., PHY packet error rates below $10^{-3}$).
Information theory tells us that this combination is fundamentally hard to
support \cite{BerGal02,PolVer10}; doing so results in low spectral efficiency,
e.g. low rate codes).  Indeed, such is the case in 5G: heavy repetition coding
is prescribed for dedicated URLLC transmissions, discouraging their use.
Meanwhile, as discussed above, other 5G advances already provide significant
latency improvement relative to 4G.  Rather than asking 6G to squeeze ever more
from the communications KPIs, we see 6G's role as actually achieving 5G's
aspirational KPIs, but with much better power and cost efficiency.

\begin{figure}
    \centering
    \includegraphics[scale=0.6]{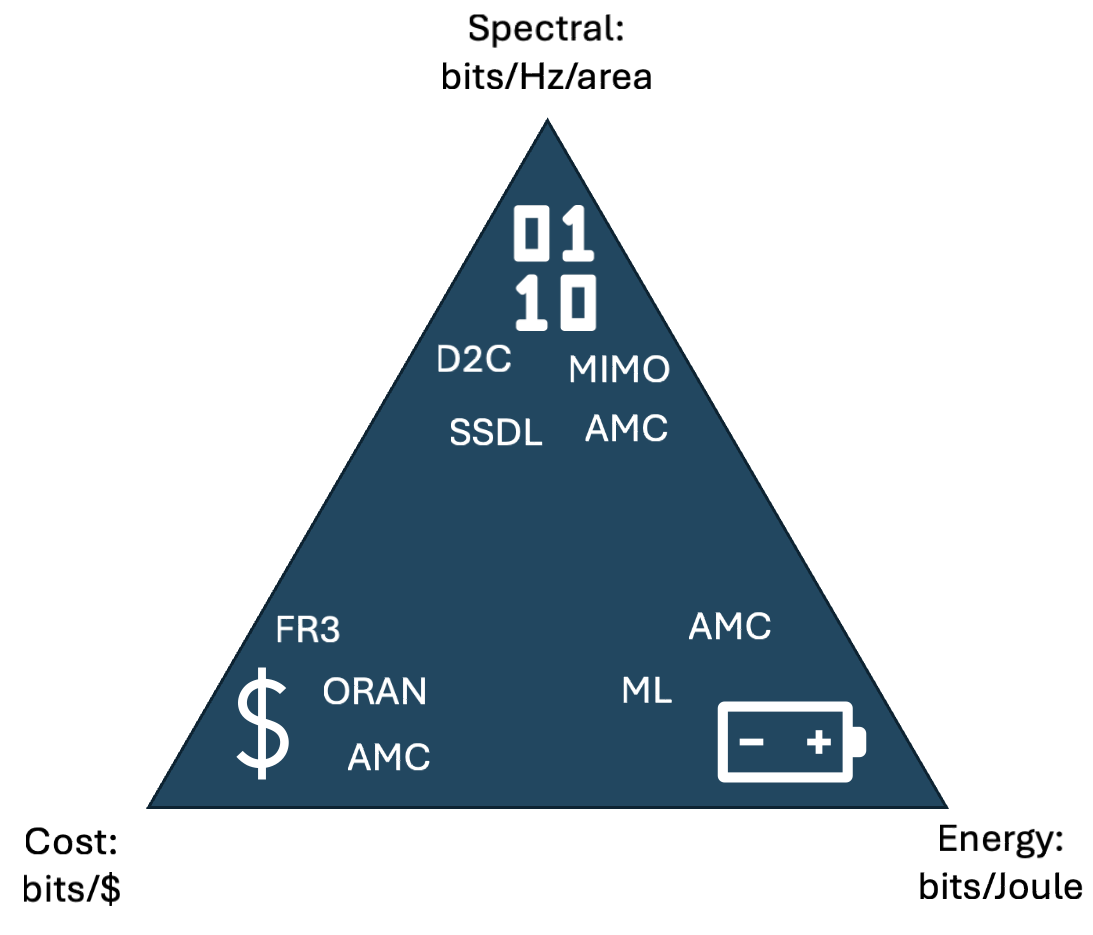}  
    \caption{6G Efficiency Triangle.  6G system design will not focus so
      strongly on maximizing spectral efficiency and delivered data rate.  Cost
      and energy efficiency will take on peer status, if not primacy, in
      communications system design.}
    \label{fig:hyper-eff-triang}
\end{figure}

\textbf{The 6G efficiency triangle.}  Rather than the 5G use-case triangle, it
is perhaps helpful to think of 6G design requirements in terms of the ``6G
efficiency triangle'' shown in Fig. \ref{fig:hyper-eff-triang}.  The three
vertices of the 6G efficiency triangle are (i) area spectral efficiency
(bits/Hz/area), (ii) cost efficiency (bits/currency unit), and (iii) energy
efficiency (bits/Joule).  Other important efficiencies such as computational
efficiency and the area efficiency (in hardware such as a chip) can be largely
projected onto these three dimensions; e.g., a lower-complexity algorithm uses
less energy and computing resources and thus costs less as well.  Although
cellular systems have always striven for energy and cost efficiency---especially
on the UE side---such considerations will take a quantum leap forward in the 6G
era and be of increasing importance in all design decisions \cite{TedWaste}.
This is because cost and energy consumption have become the primary bottlenecks
to cellular network growth and expansion.

\textbf{Order of magnitude network capacity increase.}  A more familiar next-G
aspect that still bears mentioning is that the 6G network will need to carry
about 10x the data traffic of 5G.  Cellular network traffic reliably
experiences exponential growth, both globally as well as locally (in a given
city, for example). Although the annual growth rate has decreased in recent
years to roughly 25\% a year in 2024 \cite{EricssonMobilityReport}---or only
half of that in some mature markets \cite{Orange2024}---a 25\% annual growth
still corresponds to a 10x increase over a decade, the traditional cadence
between Gs. Furthermore, even if the growth rate in downlink traffic decreases, rapidly growing applications such as on-device AI and video conferencing will drive brisk growth in uplink traffic.  Thus, it is a safe bet that 6G should be able to deliver roughly an order of magnitude higher total capacity than 5G.  Usually this comes down to a combination of (i) more
spectrum, (ii) higher spectral efficiency, and (iii) cell-splitting gains via network densification.  These three factors are roughly multiplicative,
and until 5G, network densification has contributed the most new capacity of the three
\cite{AndCla12}.  However, in 5G that has not been the case: the new C band
spectrum added has been unusually significant (see Section \ref{sec:spectrum}),
while densification has slowed due to costs (the reduction in cost of deploying a cell with a 10x smaller coverage area than a macrocell is far less than 10x vs. a macrocell).

\subsection{Question \theQnum: Besides ``fixing 5G,'' what new use cases,
  services, and applications will drive 6G?}
\refstepcounter{Qnum} 

\begin{figure}
    \centering
    \includegraphics[scale=0.6]{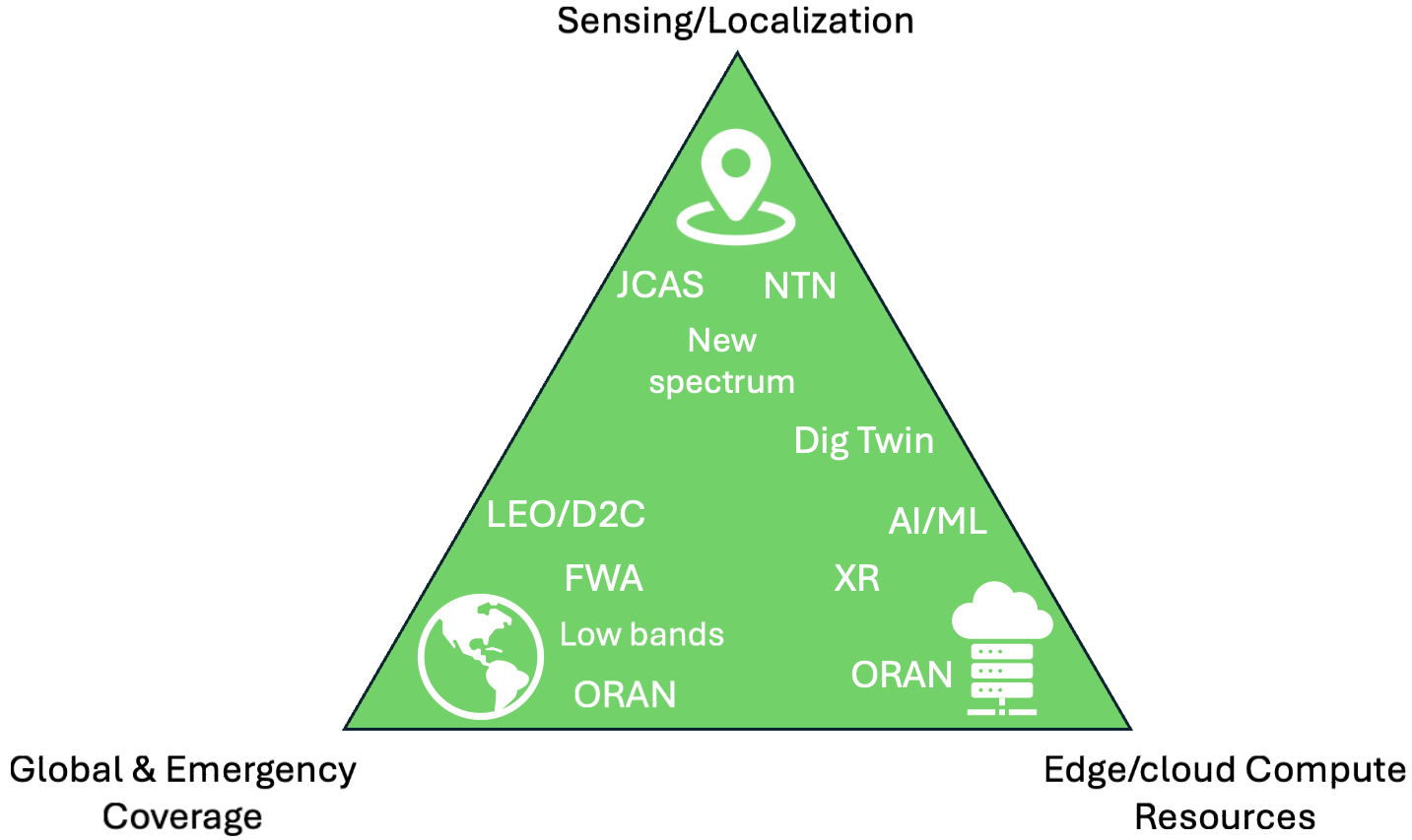}  
    \caption{6G New Services Triangle. The 6G network will provide three key new
      types of value: pervasive sensing, compute resources, and true global
      connectivity including emergency access.}
    \label{fig:Service-triang}
\end{figure}

Each successive G has faced this important but challenging
question, with answers evolving over the decade-long development of the new G.
Any answer given early in development risks miscalculation, and one might reasonably argue that until the answer is clearer for 6G than it is now, that standardization activities should wait.  Nevertheless, it is instructive to imagine what the 6G network might be tasked with in the 2030s,
and what new services it could offer.  Here, we do find the ITU-2030
framework more persuasive, which focuses on three proposed new dimensions that are the other three aspects of the ITU-2030 6G Hexagon: (i) integrated
sensing and communications, (ii) AI and communications, and (iii) ubiquitous [global]
connectivity.  In this vision, which we summarize in
Fig. \ref{fig:Service-triang}, the 6G network becomes a collection of sensors,
intelligence, and computing resources that produce, process, and consume data,
not merely transport bits. Here we discuss (i) and our take on (ii) which is more focused on compute.  We leave (iii) for
Sec. \ref{sec:Network} where we discuss global connectivity in detail.

\textbf{Pervasive Sensing and Digital Twins in 6G.} Several possible new
applications in the 6G era -- immersive XR, autonomous vehicles, mobile robotics --  will require a combination of strong situational awareness and reliable access to low-latency cloud-like computing. The 6G network---with millions of radio nodes covering ever-more of the world's population---will be uniquely situated to provide such services.  Take RF sensing.  If the 6G network
can achieve heightened situational awareness of user locations, trajectories,
and the 3D environment around them, it can improve communications functionality
such as beamforming and handover.  But the value of sensing extends much
further.  6G networks will leverage their unique footprint within urban areas to
create an urban digital twin with unprecedented spatiotemporal resolution,
starting with advanced channel estimation, then exploiting additional sensing
such as integrated OFDM-based radar \cite{sturm2011waveform}.  6G networks could
even provision low-cost on-demand-deployable drones at some BSs to monitor local
spatial dynamics (e.g., traffic patterns, construction).  Finally, the 6G
network can leverage user sensing, e.g., from XR headsets, to build continuously
updating 3D models of the urban landscape.

This new age of sensing and awareness will give rise to a unique situation: 6G
networks will have a precious asset---the digital twins of thousands of urban
areas---with application and value well beyond communications.  Firms
specializing in logistics, transportation, construction, and marketing will
subscribe to anonymized feeds from an operator's digital twin, allowing the
operator to diversify and grow its revenue.  No other entity in an urban
area---governmental or commercial---will have access to a digital twin that
could rival the one created by MNOs, because no other entity will have the killer
combination of (i) dense and vast sensing footprint, (ii) pervasive high-speed connectivity to both wireless users and the core network, and
(iii) ability to deploy and access vast computational resources, including close to the users at the network edge \cite{liu2022integrated}.

\textbf{Compute as a service.}  Connected intelligent edge with integrated AI
and communications is another compelling new service direction that will define
the 6G era \cite{JunKaibinEdge20}.  This direction will be enabled in part by merging the cellular network and cloud services. In fact, the 6G cellular network itself will become a significant cloud provider, offering common compute resources including CPUs,
GPUs, neuro processing units (NPUs), and even ASIC/FPGA accelerators to its
subscribers and to developers of apps and services
connected to the network.  One key challenge to address in this regard is the
discovery and management of compute resources on mobile devices, far edge, near
edge, and in the public cloud. Another challenge is the balancing act needed to
strike the right performance/power/latency tradeoff for edge devices.  Consider
a pair of AR glasses, for which on-device AI could provide low latency,
security, and privacy, whereas the 6G link could vastly augment the AI
capability while offering potential power savings. Given such compelling
cloud-related opportunities, we anticipate that MNOs will accelerate
transformation of cellular networks to offer a variety of computing services to
complement their communications offerings. 

\subsection{Question \theQnum: What will be the new value proposition for 6G?}
\refstepcounter{Qnum}

\begin{figure}
    \centering
    \includegraphics[scale=0.5]{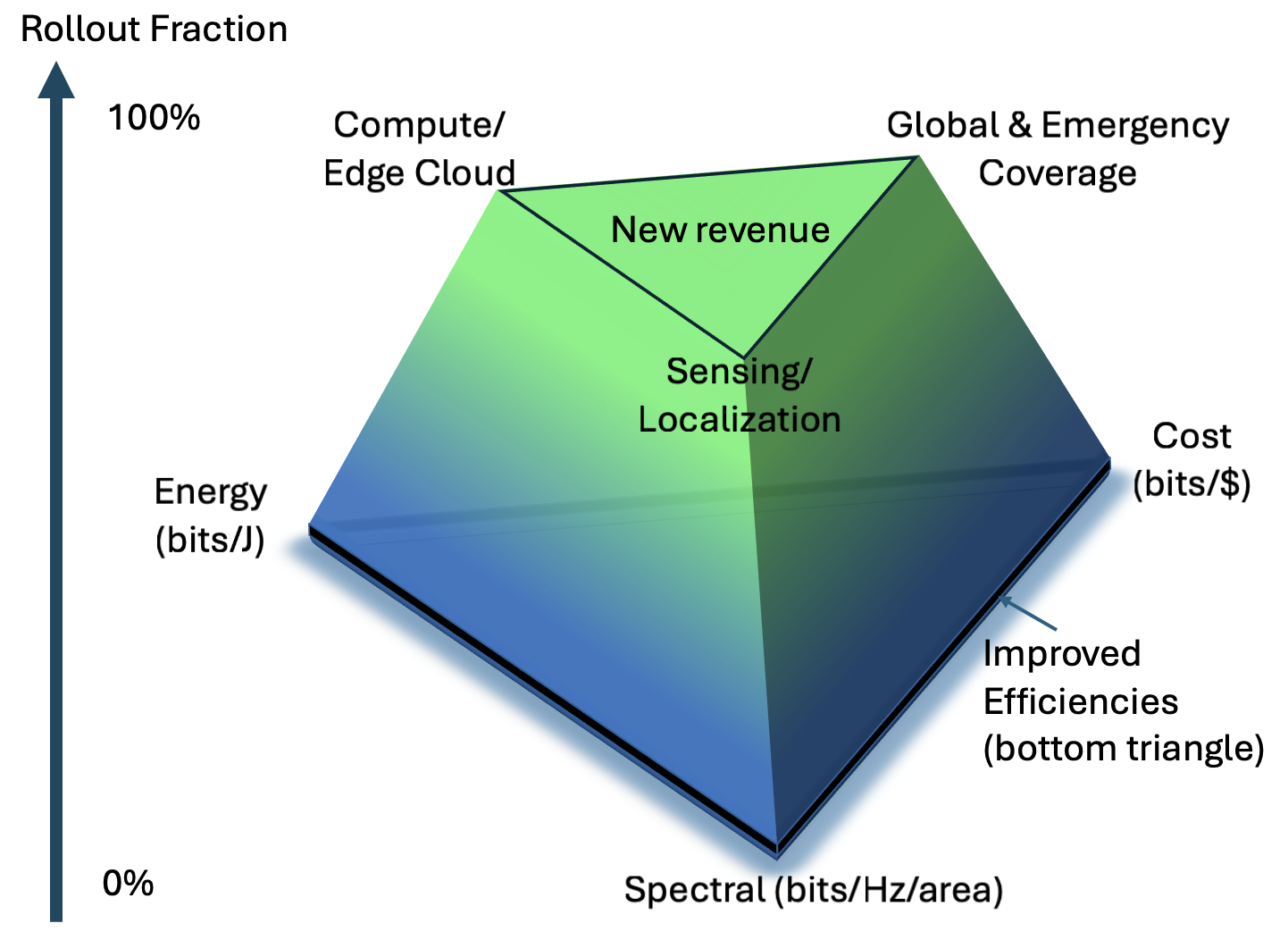}  
    \caption{6G Network Value Prism. The 6G network's newly created value versus 5G can be visualized as the volume of a prism.  The top of the prism is the 6G new services triangle, while the bottom is the efficiency triangle.  The extent of the 6G rollout is the height of the prism.}
    \label{fig:network-value-prism}
\end{figure}

The increase in value of the 6G network versus the existing 5G network can be
visualized by combining the 6G efficiency triangle and the 6G services triangle
to create a prism, as shown in Fig. \ref{fig:network-value-prism}.  

\textbf{The 6G Value Prism.} The volume of the prism is proportional to the additional value of 6G vs. 5G.  If this additional volume/value is significant, 6G will be a success.  In particular, the value of 6G will correspond to how successful the three new hypothesized services are and how widely they are adopted.  The higher the adoption of
sensing, compute, and global connectivity, the larger the area of the top
triangle and thus the volume.  Similarly, the more efficient the 6G network is in terms of energy and
expenditures vs. 5G---as well as the total amount of data rate it can
deliver---directly translates into value for MNOs and consumers in the form of
increased data consumption and lower expenses.  Finally, the state of the
rollout and market penetration of 6G directly determines the value it provides:
this is the vertical axis of the prism.  As 6G infrastructure, services and
mobile devices become pervasive -- and this will include increased fixed wireless access (FWA) -- the value of the new services and gained
efficiencies becomes more fully realized.

\textbf{Example Use Case: Mobile XR.} Consider mobile Extended Reality (XR), one of the more demanding potential mass-market use cases that could define the 6G era and provide significant new value.  If mobile immersive experiences become mainstream, XR devices will generate and consume large amounts of data, impose firm latency and reliability constraints on the network, have very limited battery life, introduce new demands for edge computing, and require new sensing capabilities: all key aspects of 6G \cite{6GXR}.  From the point of view of a 6G standard, though, the communication KPI requirements for mobile XR correspond roughly to a midpoint between
the eMBB and Mission Critical/URLLC use cases in 5G.  Indeed, 5G Advanced is
already focusing on XR as a use case, achieving latency reduction largely
through better coordination between the RAN and application interfaces, as
opposed to using or expanding URLLC modes.   

6G can definitely better support mobile XR than 5G, but less in terms of better communications KPIs, as much as in terms of delivering energy efficient, low-cost provisioning of high data rate, and providing much-improved computing and sensing resources.  Although XR is just one example 6G-era application, most others follow a similar logic.  What is needed in 6G is not the enablement of entirely new verticals and applications with improved KPIs, but rather the ability to support the imagined 5G verticals in a much more efficient way, while enabling new compute and sensing services and expanding the connectivity footprint.




\begin{table}
    \centering
    \begin{tabular}{lll}
    \toprule
        \textbf{Abbreviation} &  \textbf{Definition} &  \textbf{Meaning/Examples}\\
        \midrule
        AI/ML & Artificial Intelligence/Machine Learning & Broad terms, including Deep learning, DNN, RL \\
        AMC & Adaptive Modulation and Coding & Change data rate based on SINR \\
        AR & Augmented Reality & Overlaying digital content onto real-world environments\\
         BS & Base Station & Includes antennas, RF, baseband to at least Layer 2\\
        CSI & Channel State Information & Information on current wireless channel conditions\\
        D2C & Direct-to-Cellular & Broadband service directly between satellites and cellphones\\
        DL:UL & Downlink:Uplink  & BS to UE: UE to BS \\
        DNN & Deep Neural Network & A trainable NN with at least 3 layers\\
        eMMB & Enhanced Mobile Broadband & High rate best effort data in 5G \\
        FWA & Fixed Wireless Access & Broadband to the home via 5G/6G \\
        HAPS & High Altitude Platform Stations & Communication platform at 18-25 km altitude\\
        IoT & Internet of Things & Any automated device connected to the Internet \\
        JCAS & Joint Communication and Sensing & Using a single radio to sense or localize as well as communicate \\
        KPI & Key Performance Indicator & Data rate, latency, reliability, etc.\\
        LEO & Low-Earth Orbit &  Earth-centered orbit with an altitude of 2000 km or less\\
        MAC &  Medium Access Control Layer & Layer 2: Resource Allocation, Scheduling, Retransmissions \\
        MNO & Mobile Network Operator & Cellular operator with mobile data plans\\
        MSS & Mobile Satellite Service & Satellite communication for mobile devices and users\\
        MTC & Machine Type Communications & Communication between devices without human intervention \\
        OFDM  & Orthogonal Frequency Division Multiplexing & Becomes OFDMA at Layer 2 \\
        O-RAN & Open Radio Access Network & See Sec. \ref{sec:ORAN}  \\
        PHY & Physical Layer, Layer 1 &  Baseband signal processing operations \\
        RIS & Reflective Intelligent Surfaces & A surface optimizing wireless signal transmission\\
        RL & Reinforcement Learning & Important subclass of online AI/ML \\
        SCS & Supplemental Coverage from Space & Satellites extending mobile network coverage to remote areas\\
        SINR & Signal to Interference plus Noise Ratio & Signal quality \\
        XR & Extended Reality & Headset capable of Augmented and/or Virtual Reality\\
        NTN &  Non Terrestrial Network  & Satellites, Drones, Blimps\\
        UE & User Equipment & Mobile XR device, smartphones\\
        URLLC & Ultra reliable low latency communication &  ``Mission Critical'' services with strict real-time requirements \\ \bottomrule
    \end{tabular}
    \caption{Abbreviations and Terms in Article}
    \label{tab:terms}
\end{table}

\subsection{Question \theQnum: Will 6G be backward-compatible with 5G?}
\label{sec:backward}
\refstepcounter{Qnum} 

The goal will be to make 6G as backward compatible as possible, and to a large extent, it will be.  Paradoxically, 6G's emphasis on continuity will be one of its novel aspects, and likely to be
the norm for 7G and beyond (to the extent they are still numbered).  Regardless
of explicit backward compatibility, 6G will be designed to allow most
5G-era deployments to continue working through the 6G era, if desired.

\textbf{Continuity does not mean constrained.}. Perhaps counter-intuitively, an emphasis on continuity could speed innovation in the 6G era and beyond.  Rather than waiting for a new G for a quantum step forward, significant innovation can instead occur within a common agreed-upon framework.  New features can be added quickly without uncertainty over the baseline aspects of the standard.  We have seen this trend already in 5G, where
there is a new emphasis on ``forward compatibility,'' largely by means of
flexibility and modularity.  5G examples include bandwidth parts \cite{LinBWP},
flexible OFDM subcarrier spacing, self-contained TDD slots, and almost unlimited
carrier aggregation.  Although previous Gs did break backward compatibility,
this was largely because they changed fundamental aspects of a monolithic physical layer (e.g., 3G CDMA to 4G OFDMA), making backwards compatibility impossible.
Such clean-slate designs will no longer be necessary or desirable.

\textbf{6G's long life expectancy.} Still, a strict definition of ``backward compatibility" would be that all 5G
features have to be supported by a 6G base station.  This will almost surely not
be the case in 6G: we expect some 5G features to be dropped for 6G.  But we do
expect a fluid co-existence between 5G and 6G, and for 5G's lifetime to have a
very long tail.  Many IoT applications such as smart meters and cars can have a
lifetime of decades---10x longer than a smartphone---so having stable
deployments is indispensable for those applications.  Indeed, many IoT business
cases are not viable if they cannot count on a working cellular connection ten or even twenty years after a G transition.  Thus, we expect RedCap and other 5G IoT-centric features to be either supported explicitly by 6G, or for 5G RedCap deployments
to persist indefinitely in a portion of the 6G era spectrum, as discussed in
Sec. \ref{sec:spectrumCake}.




\subsection{Question \theQnum:  Will 6G be the last G?}
\label{sec:endofGs}
\refstepcounter{Qnum} 


Not entirely.  From both spectrum allocation---which occurs on roughly a
once-per-decade cadence---and marketing/commercial points of view, there will be
major forces keeping the new-G-per-decade tradition alive.  But from a technical
point of view, 6G will be a major step towards ending the entire xG paradigm.

\textbf{Will 6G be the ``Pentium" of cellular?}  By way of a historical analogy, consider the seminal x86 series of Intel microprocessors that ushered in the personal computing era.  After the initial version in 1978, there
was the 286 (1982), 386 (1985), then 486 (1989), before numbering fatigue
settled in and the 586 was branded the “Pentium” in 1993.  The Pentium became
the dominant Intel microprocessor branding for over a decade---spanning many
subsequent “generations”---and essentially terminated the incremental x86
numbering system.  This does not mean that innovation stopped or that
microprocessors did not evolve significantly in that time period.

\textbf{Decoupling technical innovation from standards and marketing.}   We predict that a similar shift is in store for cellular networks, perhaps as
soon as 6G.  Maybe we will see a Pentium-like decoupling of marketing campaigns for 6G era networks from the technical innovations going into them.  As we just discussed, 6G will be a more modular and upgradable network (see also Sec. \ref{sec:ORAN}), where innovations can be adopted on an arbitrary schedule, as opposed to a short-lived flurry of innovation with the introduction of each new G.  This trend will become unmistakable for the 5G to 6G transition.  This will be largely positive, as it will enable more innovation to occur between Gs rather than incentivizing an undue focus on the design and standardization of a new G, which can be quite limiting.

\section{Spectrum for 6G}
\label{sec:spectrum}

The release of new spectrum has been a defining part of each G.  And starting
with 5G, there has been a growing recognition that the characteristics of
different spectral bands are suited for different use cases.  5G brought in two
important new blocks of spectrum: the ``C band'' (3.5-4.2 GHz), and millimeter
wave (mmWave), which for 5G comprises a few GHz of spectrum in the 28-40 GHz
band. For both C band and mmWave, particular frequency allocations vary by
country.  Although less hyped in the 5G build-up, the C band has been much more
impactful than mmWave.  In fact, 5G C-band deployments have been so successful
that in many markets operators are currently experiencing an unusual phenomenon:
spectrum abundance.  Of course, as more 5G devices come online in the coming few
years, spectrum shortages will resume based on current trends.  Thus, there will
be considerable pressure to find new spectrum for the 6G cellular network to
ensure sufficient spectrum in the early 2030s.

\subsection{Question \theQnum: Which spectrum bands will have a major impact on 6G?}
\label{sec:FR3}
\refstepcounter{Qnum} 

6G will eventually subsume all the 5G spectrum and bring online new spectrum,
hopefully with a high level of global coordination.  \textbf{The most important new
spectrum for 6G will be in the so-called Frequency Range 3 (FR3), which is
defined as roughly 7-24 GHz.}  The biggest impact will likely come from the lower
portion of that band, particularly just above 7 GHz.  China may use the upper 6
GHz band (e.g., 6.4-7.125 GHz) for 6G, but in the United States (US), most of
the 6 GHz band has been allocated as an unlicensed band, so 6G services are more
likely in the 7.125 to roughly 8.4 GHz range.  At least some regions may also
allocate considerable spectrum in the 13-16 GHz range for 6G cellular, including
in the US.

The 7-8 GHz ``lower midband'' is well-suited for 6G and is essential to
achieving 6G's intended 10x capacity increase for a few key reasons: (i)
propagation in this band is acceptable for cellular coverage, and unlike mmWave,
can achieve outdoor-to-indoor penetration; (ii) with a modest scale-up of the
antenna array dimensions, but with the same panel size, a coverage area similar
to the current C-band deployments can be achieved, allowing the same cell sites
to be utilized, which is critically important for cost savings; (iii) the amount
of bandwidth available is large, roughly on par with all current sub-6 GHz
spectrum; and (iv) considerable spatial multiplexing is expected to be possible.
The last point is due to the fairly large antenna arrays that are feasible at
this smaller wavelength (about 4 centimeters)---for example, a $256 \times 16$
MIMO channel could be possible---and the expectation of sufficient scattering
in this band \cite{KangMidband24}.  For NLOS channels, a channel rank not too far below the upper
bound of $\min(N_t,N_r)$ may be achievable in this band, particularly for
MU-MIMO channels.

The 12.7-13.25 GHz band and 14.8-15.35 GHz band identified by the FCC and ITU
for potential 6G use are also intriguing.  It could end up meeting the promise
of localized ultra-high capacity in urban areas with the contiguous outdoor and
indoor coverage that mmWave has so far failed to deliver.  The band's wavelength
is short enough to permit MIMO configurations with many more antennas, yet long
enough for better coverage, less sensitivity to beam misalignment, and
more-nearly fully digital beamforming in practical low-cost antennas.  The
superiority of this band relative to mmWave as far as implementation is
demonstrated by the extremely low-cost mass-market Starlink user terminals,
which operate in the nearby $K_\text{u}$-band (12-18 GHz) frequencies.

\subsection{Question \theQnum: Will millimeter wave actually be used in 6G?}
\label{sec:mmWave}
\refstepcounter{Qnum} 

6G will make wider use of the abundant mmWave spectrum (28 GHz and above) than
5G has so far.  Still, mmWave will be primarily utilized for capacity upgrades
in very dense areas, venues like stadiums and concert halls, airports and malls,
and downtown pedestrian-centric areas.  It will also be used for Fixed Wireless Access in medium to high density areas (as it is already doing in 5G), and also be an enabler for new RF-sensing use cases, such as radar and localization.  The expectation that
widescale mmWave deployments would be the defining feature of 5G was always
dubious: the coverage area of mmWave cells is just too small, and it is
very difficult to provide outdoor-to-indoor coverage due to large penetration
losses (10s of dB). In the first half of the 5G era, baseline coverage has been
the highest priority for operators, and there has been limited interest in
deploying mmWave.  Most mmWave deployments are in the United States, with only a
few thousand sites in other countries.  For example, there is no mmWave cellular
in China, despite China having strong uptake of 5G and many densely populated
cities.  However, as 5G achieves ubiquitous coverage (primarily via C band), the
interest in mmWave deployments as a capacity multiplier will sharpen.
Furthermore, more mmWave-ready handsets will be in circulation, increasing the
motivation for such deployments, while bringing down costs due to economies of
scale, creating a virtuous cycle.

In 6G, we also expect significant enhancements to mmWave from an engineering
standpoint.  These include
\begin{itemize}
\item Increased range via improved and enlarged antenna arrays at both the BS
  and handset.
\item Faster and more precise beam alignment, enabled largely through deep
  learning methods, see Sec. \ref{sec:MLPHY2}).
\item Improved mobility support via beam tracking and prediction, again aided by
  machine learning tools and offline training with digital twins.
\item Greatly reduced power consumption, owing to holistic energy-centric design
  approach that improves the efficiency of the device itself, and to a 6G
  protocol that minimizes the time spent transmitting and receiving (e.g., in
  beam tracking and alignment).
\item Hardware improvements for both performance and cost, as economies of scale
  in mmWave have only been reached very recently.
\item Improved backhaul options for mmWave small cells, e.g., with in-band or
  out-of-band access link integrated into the mmWave small cell \cite{CudGho21}.
\end{itemize}

We expect mmWave will become a much more competitive technology by 2030.  That
it has struggled in 5G is not unexpected for such a radically new cellular
technology: MIMO also took over a generation to impact cellular systems.



\subsection{Question \theQnum: How will the different spectral bands in 6G be
 utilized?}
\label{sec:spectrumCake}
\refstepcounter{Qnum} 

\begin{figure}
    \centering
    \includegraphics[scale=0.5]{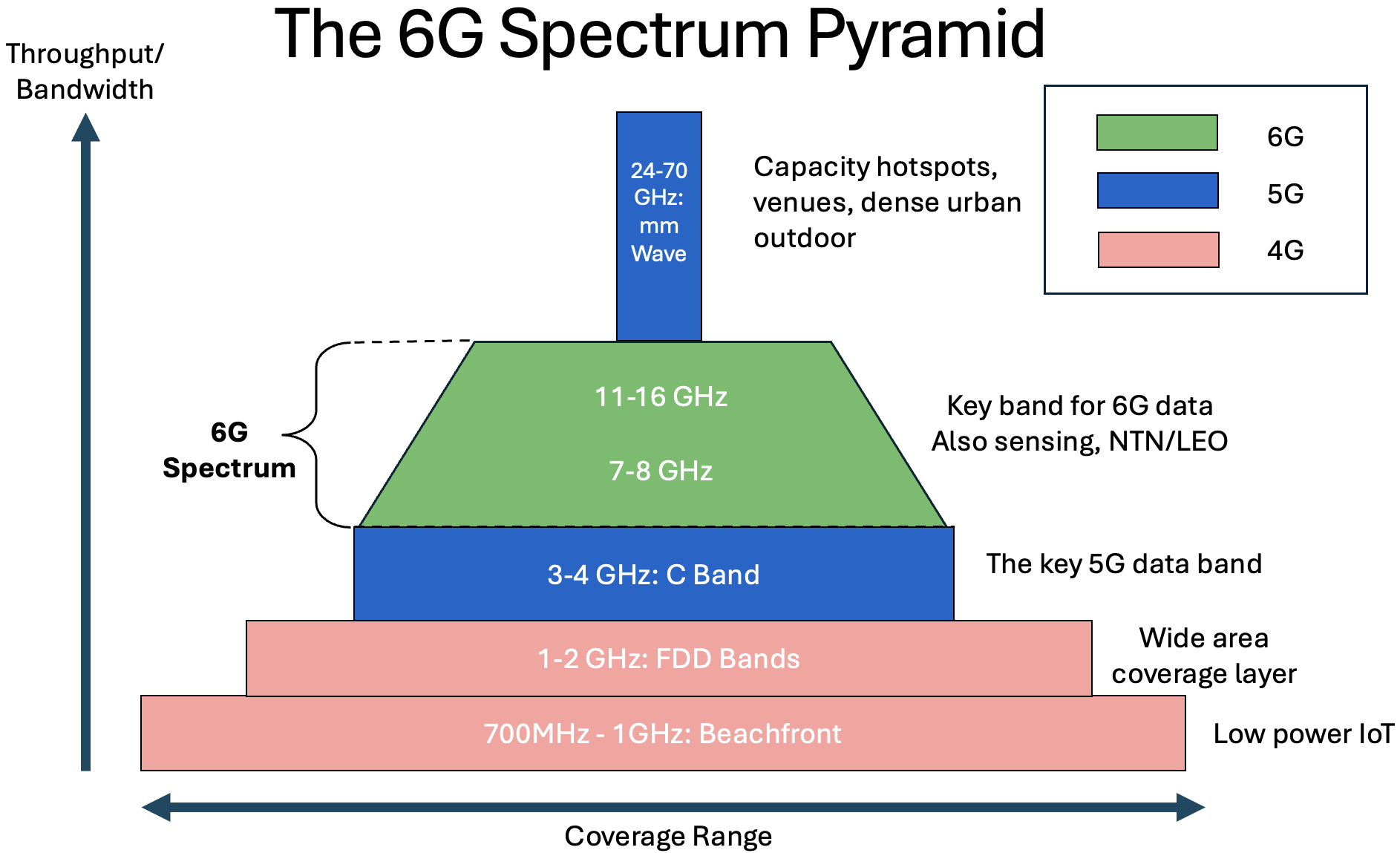}  
    \caption{6G Spectrum Pyramid.}
    \label{fig:SpectrumPyr}
\end{figure}

We envision 6G providing a sharper demarcation of spectrum for different use
cases, as summarized in Fig. \ref{fig:SpectrumPyr}, which takes inspiration from
Nokia's spectrum ``wedding cake'' \cite{HolmaNokia6G}.

\textbf{Under 1 GHz: The IoT Band.} The lowest bands, such as the 700 MHz band,
should be primarily used for wide-area, low-power, low-bandwidth applications
such as public safety, voice and IoT services, including metering and tracking.  Smartphone and video
traffic should be moved almost entirely off these bands, except in rural or
other coverage-limited scenarios where they can serve as a fallback.

\textbf{1-2.5 GHz: The 6G FDD Coverage Layer.}  The FDD paired spectrum in the
range of roughly 1-2.5 GHz should be used as the wide area coverage layer for 6G
UEs. Control information and high priority data can particularly benefit from
the robust propagation in these bands.  Because FDD bands are paired, the uplink
bandwidth is equal to the downlink, and thus comprises a valuable and relatively
plentiful resource for the transmissions of power-constrained UEs. Furthermore,
we believe it is possible to get a 70-100\% capacity gain even in these bands
from the combination of several baseband physical-layer innovations such as
improved MIMO feedback and precoding, improved coding and modulation including
in band duplexing (see Section \ref{sec:AMC}), reduced frequency guard bands,
and improved reference symbol design and utilization.  Even larger gains can be
achieved if the antenna arrays are upgraded to support massive MIMO.  In short,
this conventional band will support wide area fallback coverage for 6G
smartphones and data services.

\textbf{3-4 GHz and 7-16 GHz: The 6G Capacity Workhorse.}  The C band, which has
been by far the most successful new deployment for 5G, will continue to be an
important workhorse for 6G and may broaden to include spectrum in the 3.1-3.4
GHz band. High speed data traffic will be primarily carried by this band and the
FR3 band (7-16 GHz). Collectively, these bands will provide a continuum of
capacity-coverage tradeoffs and will carry the vast majority of 6G bits.
Although operators should be able to reuse nearly all C band cell sites for FR3
base stations, the C band radios and physical layer in 5G cannot simply be
scaled up to FR3 for 6G.  Even at the lower end of FR3 such as around 7 or 8
GHz, the number of antennas per unit area is roughly 4x that of C band.
Although increasing the number of antennas does not typically increase the total
power, since the per-antenna power is scaled down, we expect a more
sophisticated hybrid architecture will be necessary at 7 or 16 GHz to
efficiently exploit the additional antenna elements and maximize capacity (MIMO)
and coverage (antenna gain).  Nevertheless, RF integrated circuit technology
supporting frequencies up to about 16 GHz is quite advanced, and can provide
low-power amplification, unlike at mmWave.  Thus, macrocell BSs providing up to
1 Gbps at a range of 1 km are possible even in the 13-16 GHz band.  We group
these bands together to emphasize their continuum in providing high capacity
with gradually decreasing coverage.

\textbf{mmWave: Capacity Hotspots.}  The mmWave band offers routes to localized
high capacity links and will become more widespread in 6G than thus far in
5G. But their use will still be focused on urban and venue hotspot applications, and also fixed wireless access; as discussed in Section \ref{sec:mmWave}.  Nevertheless, well-positioned hotspots can provide considerable traffic offloading from lower bands and macrocells, raising network performance as a whole.

\subsection{Question \theQnum: You forgot to mention Terahertz (THz) spectrum
  and Reflective Intelligent Surfaces (RIS), which are often touted to be a big
  part of 6G.  What’s their role?}
\label{sec:THz}
\refstepcounter{Qnum} 

We do not envision a major role for THz (colloquially, carrier frequencies above 100 GHz) or RIS in 6G as regards providing significant coverage or capacity enhancements.  The mmWave experience so far in 5G does not lend optimism to use of the THz bands---loosely meaning spectrum
above 100 GHz \cite{TedTHz19}---for terrestrial cellular deployments in the
near future.  Currently, several of the major service providers we are aware of are carrying well below 1\% of their 5G data traffic on mmWave bands, driving home the point that a successful cellular technology must deliver reasonably wide area coverage and penetrate sufficiently to provide service indoors as well.

THz could, however, be useful for a few other purposes in the 6G era.
These include (i) short-range sensing such as localization and gesture
recognition, where the large bandwidth and short wavelength of THz bands are
well-suited for producing high-resolution images; (ii) terrestrial
point-to-point LOS outdoor backhaul; (iii) indoor short-range data-center
interconnects, which will be increasingly high speed in the 6G era; and (iv)
non-terrestrial network (NTN) and LEO applications, including inter-satellite
communication (see Sec. \ref{sec:Network}).  The main innovations needed for THz
communications are largely in the circuit and hardware aspects, e.g., making
the large phased arrays power efficient, dynamically steerable, and affordable
(mass-producible).  There will also be a strict need to ensure that the many
ultra-sensitive sensing incumbents in the THz bands, including weather
satellites and radio astronomy, are not adversely affected by terrestrial
communications in adjacent bands \cite{TedTHz19}.

RIS are an interesting recent development from a technical point of view
\cite{BasRIS19,RISSurvey24}, and can in principle significantly extend the
coverage range for mmWave and THz communications, particularly for NLOS
scenarios, even solving the outdoor-to-indoor penetration problem when the RIS
is integrated into a window.  But the cost of widescale deployment and
configuration of RIS remains a challenge, particularly when the benefit-vs-cost
of deploying a RIS is compared to that of deploying a full-fledged 6G small
cell.  

\subsection{Question \theQnum: How will spectrum be allocated in
  6G?  Will dynamic spectrum sharing across operators and systems become the
  norm?}
\label{sec:DSS}
\refstepcounter{Qnum} 

6G will necessarily require novel and more varied spectrum allocation models
than previous Gs.  Starting with FR3, which as we argue above is the key new 6G
spectrum, there are numerous incumbents in this spectrum, including a variety of
fixed and mobile satellite users and also defense-related users \cite{Gho24}.
While identifying and protecting such incumbents is in principle possible, it
will probably require a more sophisticated approach to spectrum management than
has been achieved to date in cellular networks.

\textbf{Spectrum sharing in 6G will proceed cautiously.}  However, we emphasize
that complicated and restrictive spectrum management could significantly reduce
the commercial value of any spectrum to mobile operators in two aspects.  First,
service outages due to pre-emption by incumbents are not acceptable for cellular
customers.  Thus, if spectrum cannot be used reliably, it is of little value to
operators.  Second, to the extent that spectrum sensing adds complexity, power
consumption (from constant monitoring), or complicated network operations, this
will encounter strong resistance since it runs contrary to 6G's value
proposition, which is predicated on cost and power reduction.  We expect that a
successful and commercially viable 6G spectrum strategy will involve a mix of
relocation of incumbent services and sharing between the incumbents and 6G
services on a small fraction of the total available spectrum to minimize the
impact on 6G service quality.


That said, an overall theme for 6G will be that the era of ``clean'' spectrum
auctions is probably over: nearly all \emph{new} spectrum will come with various
caveats, and at least modest forms of spectrum sharing will become more typical.
One can look to the sky for an early example of forced spectrum sharing, with
the emerging dense LEO constellations such as Starlink, OneWeb, and Kuiper being
required to coexist in the same spectrum under complicated and still-evolving
rules.


\textbf{Sharing in 6G is not just about spectrum sharing.}  On the topic of
sharing in 6G, network equipment sharing is likely to be even more impactful and
widespread than spectrum sharing.  Network operators in many countries already
widely share BS towers.  For example, a typical macrocell tower in the US might
be owned by a third party such as Crown Castle or American Tower, and house the
(individual) antennas, radios, and possibly other hardware for two or three
operators.  This trend will accelerate in 6G to include the rest of the
hardware, with the aim of achieving higher cost and energy efficiencies.  In
China, government-mandated infrastructure sharing among operators has led to multi-billion dollars of savings per year, resulting not only in lower costs
for consumers, but also in new investments such as network densification, which
have improved the consumer experience.  Given its clear benefits, expanded
network sharing is likely to be adopted voluntarily in the 6G era by MNOs in other
countries.

\section{The 6G Physical Layer}
\label{sec:PHY}
\subsection{Question \theQnum: Will there be a new waveform for 6G beyond OFDM/OFDMA?}
\label{sec:OFDM}
\refstepcounter{Qnum} 

The dominant waveform in 6G will remain scalable OFDM.  The OFDM waveform is
nearly optimum for cellular data communications in several respects: (i) it is
composed of parallel eigenfunctions for linear channels, i.e., narrowband complex sinusoids; (ii) it allows a clean separation of spatial (multi-antenna)
and intersymbol interference suppression, making it very advantageous for MIMO precoding and equalization;
(iii) it has fundamentally low implementation complexity due to the FFT
algorithm and subsequent one-tap equalization; (iv) it allows fully flexible
multiple-access in both the time and frequency domain, with each connection
allowed its a customized dynamic bandwidth; and (v) OFDMA provides flexibility to
Layer 2 and above, enabling opportunistic scheduling of orthogonal resources,
which can in principle achieve the Shannon capacity of a fading downlink or
uplink channel with full CSI \cite{TseVisBook}.

OFDM technology is also mature and ubiquitous, and for the most part beyond patent
protections, making it difficult to beat on cost.  Most of OFDM's alleged
shortcomings such as its peak-to-average power ratio (PAPR), long symbol time
\cite{MicFet14}, and slow frequency rolloff are largely overstated, and can be
addressed with a variety of well-known approaches.  Consider that the recently
designed clean-slate broadband LEO communications systems Starlink and Kuiper
have both adopted OFDM \cite{humphreys2023starlinkSignalStructure}.  This is
despite the PAPR issue, which is a more significant liability for satellite
systems than terrestrial ones due to challenging link budgets.  OFDM is
entrenched firmly in WiFi as well, and shows increasing promise for joint
sensing and communications applications
\cite{graff2024purposeful,graff2024data}.


\textbf{OFDM is not exclusive.}  OFDMA is a flexible framework allowing for
sub-waveforms, including embedded single-carrier waveforms.  By nulling some of
the subcarriers in the OFDM waveform, a non-OFDM narrowband waveform can be
transmitted simultaneously in the now-unoccupied subcarrier slots.  Even a
wideband non-OFDM signal can be multiplexed into an OFDM waveform, provided it
is sent over the entire OFDM symbol time.  For example, new sensing-centric or
estimation-centric signals that are not technically OFDM could be sent in such a
manner \cite{liu2022integrated}.  Alternatively, a meta-waveform operating on
top of OFDM, such as OTFS \cite{OTFS-BITS22}, can also be envisioned for special cases such as ultra-high mobility. Furthermore, other custom narrowband signals for IoT or other low-power coverage-limited applications could be developed for 6G that have low PAPR or high resiliency to hardware imperfections for use cases in which spectral efficiency is not the primary concern \cite{LozRan23}.  These too can be fit
into the OFDMA framework by allocating a portion of the band.

In summary, OFDM and OFDMA appear to be unbeatable in several different dimensions:
near-optimality from a communication theory standpoint; flexibility and
adaptability for different use cases; mature and efficient implementations.
Thus we posit that OFDM/OFDMA is here to stay for 6G, although it may also incorporate other waveforms for specialized applications.

\subsection{Question \theQnum: Will there be new types of coding, modulation, or duplexing?}
\label{sec:AMC}
\refstepcounter{Qnum} 


The two main desiderata for 6G coding and modulation are (i) facilitating more
cost- and energy-efficient implementations in the higher bands, and (ii)
improving spectral efficiency in the lower bands, e.g., the FDD bands discussed
in Sec. \ref{sec:spectrumCake}.

Consider developments in 5G NR in pursuit of these goals. Data channel decoding
complexity and control channel reliability were both significantly improved
compared to 4G LTE. Although the 5G LDPC code coupled with a systematic bit
priority mapping (SBPM) interleaver has shown about 0.5 dB gain over the LTE
Turbo code, the primary benefit of LDPC is its enabling a much higher area
efficiency (Gbps/mm$^2$). The inherent parallelism of the LDPC decoder and fewer
parity checks at a high coding rate lead to more than 5x gain in throughput at
peak rate for the same hardware area compared to a Turbo code
\cite{Richardson2018LDPC}.  On the control channel, the introduction of a Polar
code \cite{Arikan2009Polar} enabled higher control channel reliability at small
block length compared to the LTE tail-biting convolutional code (TBCC), while
achieving reasonable encoding and decoding latency \cite{Bioglio2021Polar}. 5G
NR also introduced a new modulation scheme, $\pi/2$ BPSK, to enable high PA
efficiency in low-rate uplink transmissions.

\textbf{Coding and Modulation will focus on cost and energy vertices of the 6G
  efficiency triangle}.  6G will explore further spectral efficiency
improvements in coding and modulation, but these will be strongly conditioned on
cost, area, and power efficiency.  It is highly likely that 6G will be required
to maintain a hardware ``component-wise" backward compatibility for 5G and 6G
hardware sharing.  On the coding side, 6G is likely to use upgraded LDPC and
Polar codes that allow the same hardware units to decode both 5G and 6G packets,
as opposed to new coding families such as rate-less spinal codes, staircase
codes, and polarization adjusted convolutional (PAC) codes, which may offer
modest performance gains in some scenarios but are hardware-incompatible with
5G.  Similarly for 6G modulation schemes, the industry will prefer new
modulation schemes that build upon QAM and can be efficiently demapped over MIMO
channels at peak data
rates. 

\textbf{Constellation shaping} is one area that could bring significant performance gains
in 6G since the 5G LDPC code is already within 1 dB of the constrained capacity
of the QAM constellation. There are two main approaches to achieving the shaping
gain, probablistic amplitude shaping (PAS) \cite{Bocherer2015PAS} and geometric
shaping, both of which could potentially lead to coded modulation performance
beyond the constrained capacity. Geometric shaping, which directly modifies the
constellation points to mimic a Gaussian distribution, has been adopted in DVB
and ATSC standards for broadcasting.  But the demapping, especially for MIMO,
may be too complex for 6G.  PAS modifies the distribution of the input bits to
an FEC followed by a QAM modulator, where the probabilistic distribution of the
constellation points approaches that of a Gaussian distribution. This is optimal
from an information theory/entropy perspective.  A prefix-code based PAS scheme
has been adopted in WiFi-7, but the distribution matcher is likely to be
redesigned in conjunction with a 6G LDPC code to meet the rate, block length,
and performance requirements.

\textbf{Duplexing in 6G.} All previous Gs have been designed for
frequency-domain duplex (FDD) and time-domain duplex (TDD).  Could 6G be the
first to embrace full-duplex (FD) \cite{Everett2014FullDuplex}?  We believe so,
for two reasons.  The first is the need for better uplink (UL) coverage and
capacity in 6G to accommodate the rise of UL traffic due to interactive
on-device applications such as XR. TDD bands offer ample downlink capacity,
assuming a typical 70-80:30-20\% DL:UL duty cycle.  But this leads to a
significant reduction of UL coverage compared to FDD, since the UL duty cycle
cuts the available power by 70-80\% versus FDD, where the UE can transmit
continuously with a smaller bandwidth.  Sub-band full-duplex (SBFD) within a TDD
channel allows a macro base station to simultaneously transmit and receive at
full duty cycle via frequency and spatial separation, using self-interference
mitigation techniques \cite{Abdelghaffar2024SBFD}.  SBFD essentially turns
wideband TDD into wideband FDD, substantially improving both the UL link budget
and capacity.  The second reason for the rise of full duplex in 6G is joint
communications and sensing (JCAS).  Full-duplex transceivers will turn the 6G
network into a robust and dense sensor network that generates valuable data
analytics for services such as security, eHealth monitoring, autonomous cars,
drones, and robotics.

\subsection{Question \theQnum: Will we see any new types of MIMO in 6G?}
\label{sec:MIMO}
\refstepcounter{Qnum} 

There are few technical topics where information theory has had a larger impact
than on multi-antenna communications (MIMO), with $N_t$ transmit antennas and
$N_r$ receive antennas.  From unveiling the promise of a multiplicative capacity
increase on the order of $\min(N_t,N_r)$ for single user MIMO in both
closed-loop \cite{Tel99} and open-loop setups \cite{FosGan98}, to showing that
such gains extend to more general multiuser MIMO scenarios
\cite{CaiSha03,VisJin03}, the now vast body of strong theoretical results on
MIMO communication is a bedrock of communication and information theory
\cite{HeaLozBook}.  However, since nearly all these results are based on a
simple narrowband abstraction of the MIMO matrix channel $\bH$, and generally
assumes perfect knowledge of $\bH$---particularly at the receiver but also
often at the transmitter---it is not surprising that the gap between theory and
practice for MIMO is one of the larger such gaps in modern cellular systems.
This is particularly true for large-dimensional MIMO systems with nontrivial
mobility, but also for low-rank mmWave MIMO channels that rely on accurate
transmit-receive beam alignment.

Before predicting the form that MIMO will take in 6G, it is useful to consider
how MIMO was rolled out in LTE, and based on lessons learned in the 4G era,
subsequently simplified and enhanced for 5G.

\textbf{MIMO in LTE: An Era of Experimentation.}  The LTE standard pondered a
wide variety of MIMO techniques, codified by eventually 10 different
``transmission modes'': TM1 to TM10\footnote{We will focus our MIMO discussion
  on data transmissions as opposed to control channels, which also make use of
  MIMO but generally in a more simple and robust form.}.  These scenarios
covered nearly the whole MIMO kitchen sink:

\begin{itemize}
\item Alamouti codes \cite{Ala98} and frequency switched diversity (TM2).
\item Open-loop spatial multiplexing with cyclic delay diversity (TM3),
  conceptually similar to BLAST \cite{Fos96}.
\item Codebook-based closed loop spatial multiplexing (TM4), following
  \cite{Lov03b} and related approaches.
\item Multiuser MIMO (TM5), which was also codebook based.
\item A suite of closed-loop approaches that could utilize arbitrary precoders
  (TM7-9) unknown to the UE side.  These were well-suited to TDD bands in
  general and massive MIMO approaches such as FD-MIMO \cite{Nam13} in
  particular.
\end{itemize}

By the peak of the LTE era, the dominant implementations comprised TM3, TM4 and
TM8 \cite{Chen5GBook}.  In particular, TM3 and TM4 were usually used for FDD
bands, while TM8 was the preferred mode for TDD.  The most common
high-performance receiver design, for all three modes, was a variant of a sphere
decoder, which offers superior performance compared to MMSE receivers in many
channel conditions.  In short, the cellular industry implicitly converged on a
few preferred approaches for MIMO in LTE, which provided several useful lessons
for 5G and 6G systems.

\textbf{MIMO in 5G: Beamforming-centric MIMO.}  5G picked up where LTE left off,
but focused further on the approach of TM7-9, which was to foresake codebooks or
open-loop MIMO and allow for arbitrary precoders that are transparent to the
UE. In essence, the 5G MIMO approach is centered around beamforming, whether for
a single strong beam at low SNR or a low-rank channel, or for multiple beams at
higher SNR and channel rank, consistent with theoretical principles
\cite{HeaLozBook}. This beam-centric approach to precoding was buttressed with
channel-reciprocity-based feedback using sounding reference signal in TDD
deployments. When the mobile sounding (pilot) signal is too weak at the cell edge, 5G DL precoding is able to use a rich feedback known as Type II CSI, which allows up to
4 strong beam directions to be fed back using linear combinations of beams
from an oversampled DFT codebook \cite{Chen5GBook}.  However, at the current
time, all commercial 5G systems still use only Type I CSI, to the best of our
knowledge.  Furthermore, MU-MIMO has thus far had a fairly minor impact, related to challenges of grouping users amenable to joint precoding (e.g. similar pathloss, current CSI available for the same subband, quasi-orthogonal channels, data available to send).  

\textbf{MIMO in 6G: Native MIMO and FDD MIMO Revisted.}  The entire 6G PHY
should be thought of in terms of being native MIMO, with all aspects built from
a MIMO-first standpoint, including pilot symbols, channel estimation, feedback,
scheduling, and the assumption of beamforming (directional transmissions).  The
main areas where cellular MIMO is still particularly suboptimal today are (i) two-directional beam alignment, for example at mmWave, and (ii) channel estimation/feedback and precoding for large MIMO channels, since it is impossible to know the entire spatial channel (with frequency selectivity) in all but the most static
environments.  With MU-MIMO, problem (ii) is quite challenging, since the
precoder depends on which UEs are scheduled. On the plus side, continued network densification and the nature of FR3 should favor MU-MIMO in 6G vs. 5G.   We discuss point (i) in Sec. \ref{sec:mmWave}.  Regarding point (ii), 6G will see a continued evolution towards more digital beamforming and more advanced hybrid precoding
architectures, and the leveraging of ML techniques to provide strong priors on
the channel estimates, as well as their quantization and compression for
feedback.  We also note that FWA could be a particularly ripe use case for MU-MIMO in 6G, given the more static channels and an often large and consistent level of downlink traffic (e.g. streaming HD video to the home).

In 5G, the FDD bands were to a large extent ignored, with all the attention on
C-band TDD and mmWave. As 6G strives to deliver new efficiencies, there will be strong renewed emphasis on FDD MIMO optimization. This will include new active antenna arrays to get much better coverage and spectral efficiency from operators' precious FDD spectrum assets.

\textbf{Stretch goals for 6G MIMO}. We also expect to see early manifestations of
“near field" and possibly ``holographic'' MIMO.  Near field MIMO occurs when the aperture of the array is large relative to the
wavelength, and renders increased multiplexing possible even in LOS conditions
\cite{RamBjo23}.  Holographic MIMO refers to ultra-dense arrays or even continuously-radiating surfaces \cite{GanMIMO24}.   For low-mobility dense urban scenarios, we will see
incremental progress towards forms of “cell-free MIMO,” with multi-BS
connectivity for increased area spectral efficiency.  Cell-free MIMO in basic
forms has been supported in 4G (``CoMP") and 5G (``Multi-TRP''), but will be
better enabled in the 6G era by a combination of reduced network latency and
additional computational abilities near the network edge \cite{Hien24}.   There will also be progress towards better understanding and exploiting the electromagnetic properties of large MIMO channels \cite{Bjo24} and how to mitigate the impact of suboptimal low power circuits \cite{MezHea24}. 

\subsection{Question \theQnum: Will the 6G PHY be replaced by neural networks and run on GPUs?}
\label{sec:MLPHY1}
\refstepcounter{Qnum} 

The integration of deep learning (DL) methods into the wireless PHY is one of
the most interesting and potentially disruptive developments in communication
systems in the last decade \cite{OSheaHoydis17,LiDLOverview19}.  In principle,
neural networks can perform a superset of the current signal processing
operations found in typical wireless transceivers. With extensive data and
training, they hold out the promise of end-to-end optimized designs that could
outperform the current model-based approaches grounded in information and
communication theory, which tend to assume particular channel and noise models,
linear or near-ideal electronics, and accurate channel knowledge at both the
receiver and to a lesser extent the transmitter.  5G-era approaches often assume
other implicit idealizations or a particular decomposition of functionality.
For example, the separation of channel estimation from subsequent equalization
or precoding may be suboptimal in many settings
\cite{HonDeepRx21,zhang2020deepwiphy}.

On the implementation side, a DL-based PHY could leverage rapid advances in deep
learning models and their parallel computation via GPUs, providing a viable
pathway to the softwarization of the computationally complex physical
layer. Fine-grained customization and frequent software upgrades of the baseband
PHY functionality become viable.  This is essentially NVIDIA's stated vision for
6G, and they have developed tools such as Siona and AerialSIM to advance this
vision.  It is an intriguing possibility, and has garnered considerable research
and many interesting designs in both industry and academia.  The reality is
likely to be more nuanced.

\textbf{We believe the 5G PHY will not be wholesale or even significantly
  replaced in 6G by neural networks}, for two main reasons.

First, at the link (point-to-point) level and focused on communication KPIs like
spectral efficiency and reliability, much of the 5G PHY is near-optimal in terms
of communication and information theory, and general enough to handle the
aforementioned nonidealities with at most minor degradation.  This is
particularly true for OFDM and frequency-domain equalization, as discussed in
Sec. \ref{sec:OFDM}, and for the LDPCs.  Although several studies such as
\cite{HonDeepRx21,DornerBrink18} have shown competitive or even moderately
improved performance from DNN-based solutions in some specific scenarios --
which is impressive---the gains are typically small (a dB or less), and largely
unproven in terms of their generality or explainability.

Second, as we have postulated, 6G will be highly optimized for energy
efficiency.  There exist very efficient implementations for the most complicated
parts of the transceiver, including OFDM (IFFT/FFT) and the error correction
decoders.  These will not be beatable by DL approaches in a complexity sense; on
the contrary, DL approaches will consume considerably more power and area.  This
is particularly crucial at the UE side.  There is a better case for a DL-based
PHY at the BS side, but it will still need to be power efficient and
trainable.  

\textbf{Nevertheless, we foresee a very important role for DL in the PHY,
just not a wholesale replacement of the existing PHY pipeline}.

\subsection{Question \theQnum: What then are the key roles for machine learning in 6G?}
\label{sec:MLPHY2}
\refstepcounter{Qnum} 

A key advantage of machine learning (ML) is its ability to address
high-dimensional problems that are intractable with classical solutions. We
foresee significant opportunities for ML to make an important impact on 6G in three broad areas.  

\textbf{Where models fail, or optimal approaches are unachievable.}. A first and
important example includes large MIMO systems.  Although the theory of MIMO is
fairly mature, as discussed in Sec. \ref{sec:MIMO}, optimal approaches are not
achievable---or even close to achievable---for 5G or envisioned 6G MIMO systems
since the number of channel dimensions in the spatial, frequency, and user
domains is so large.  Moreover, wideband multiuser massive MIMO systems must
operate with low complexity and minimal CSI.  This is a highly suboptimal
setting, and one where deep learning techniques can provide significant
benefits.  At a high level, deep learning approaches can learn to extrapolate or
predict CSI, find approximate solutions to intractable problems, and generally
exploit historical data to make near-optimal inferences.  Concrete examples
include (i) channel state information (CSI) estimation
\cite{Bal2020DosJalDimAnd, ArvTam23,AttentionMIMO22}; (ii) CSI compression and
feedback \cite{CsiNetPlus,MasGunduz21}; (iii) construction of downlink precoders
with sparse CSI \cite{SohYu21, HuDing21, ParkAnd2024}; (iv) user scheduling in
MU-MIMO \cite{ShenJun21,ChuJia23}; and (v) MIMO receivers, including joint
design of channel estimation, spatial equalization, and symbol detection
\cite{HonDeepRx21}.  

A second example is overcoming hardware non-idealities such such as low
resolution quantization \cite{BalAnd19}, or power amplifier and RF distortions
\cite{ZhaStuder20}.  These degradations are hard to model and predict, and can be unique to each type of device.   ML techniques can help preserve a desired performance level while enabling use of cheaper and lower power hardware components.  

A third example is localization.  Classical techniques based on time of arrival
(ToA) of signals exchanged between the BS and UE have difficulty dealing with
the myriad ways in which non-line-of-sight and multipath conditions manifest in
urban areas, preventing network localization solutions from reaching the
sub-meter-level accuracy targeted in 3GPP's Release 17 \cite{campolo20245gnss}.
Machine learning techniques seem well-suited to harnessing operators' vast trove
of ToA data to learn fine-grained patterns of urban blockage and multipath
without the need to develop highly accurate 3D models and perform expensive ray
tracing \cite{tedeschini2023cooperative}.

\textbf{Cell-specific learning and optimization.}  5G base stations allow a
large number of design options and configurable parameters---bandwidths,
subcarrier spacings, MIMO configurations, sectorized beam patterns, handover
thresholds, and many more---and the parameter space will only increase for 6G.
Optimal settings for these parameters vary significantly from cell to cell due
to the propagation environment, user locations, density, mobility, and the
amount and type of digital traffic being carried over the channel. These
settings should be tuned accordingly, which is a complicated problem since the
settings in one cell affect the neighboring cells.  Currently, BS settings are
configured at best heuristically \cite{PatGeMah-MLNextG24-CIPAT}.  ML approaches
can be used to automate and optimize these settings, whether through offline
large scale simulations, or online adjustments.  Our research shows that large
gains are possible: as much as 5-10 dB in terms of cell edge user SINR
\cite{TekAnd23}.  We note that online parameter tuning does not necessarily
require explicit standards support, but it does require large-scale data
acquisition.  An active role by network operators would be helpful; an
alternative is to emulate and tune a large cellular network using emerging
digital twins.

A physical-layer example is beamforming and beam alignment at higher carrier frequencies including mmWave and soon FR3.  This is a notoriously difficult problem and one that causes a great deal of latency, overhead, and power consumption in 5G mmWave systems.  By developing a sophisticated DL model of the cell-specific channel and user distribution, beam alignment can be achieved much faster and with far fewer searches \cite{HenAndAhmed24}, possibly by several orders of magnitude.

\textbf{Deep Generative Models (DGMs).}  Generative AI, the most advanced form
of modern deep learning, also holds potential for complex tasks in 6G.  Because
of their computational complexity and power consumption, we do not foresee a
role for DGMs in the real-time pipeline, but they can play an important role
offline and in the cloud.  As a first example, DGMs can be used to learn and then generate cell-specific channel distributions as required for training the aforementioned cell-specific learning approaches.  Since rich channel data are scarce, noisy,
and expensive to attain in wireless systems, one can instead train a DGM with
limited data samples and then use the DGM to generate unlimited realistic data
with the desired characteristics, which can be used for training
\cite{DosGup22,Sol20, LeePar24}.

As a second example, Generative AI may enable radical new forms of compression
for video or XR streaming applications that require only an accurate sensory
representation at the receiver \cite{Gunduz23,ParkBITS25}.  Reference
\cite{Weissman24} argues it may be possible to distill an image (or video) to
just a handful of words or other semantic objects that can be compressed to just
a few hundred bits, with modern text-to-image generative AI tools recreating an
approximation of the image at the receiver.  If XR becomes a major use case in
6G, the rendering of such content could lean heavily on Generative AI
technologies to meet its challenging bandwidth and latency constraints.  But of
course such on-device image or video creation would only be viable if reasonably
energy efficient.

\section{The 6G Network: Achieving True Global Connectivity}
\label{sec:Network}
The 6G-era network---beyond the radio interface---will evolve in significant
ways, some of which are already taking shape while others remain only vaguely
adumbrated.  As mentioned in connection with the 6G New Services Triangle in
Fig. \ref{fig:Service-triang}, we expect 6G to pursue global broadband
connectivity, including into developing countries and sparsely populated
regions.  While we focus on mobile connectivity, FWA will also be an important use case of 6G and promote global broadband connectivity in the home and office.  True global connectivity is as much an economic challenge as it is technical, with the two aspects interdependent.  For example, new technologies such as dense LEO satellite
constellations and O-RAN can make vastly expanded high-speed coverage much more
commercially viable.

\subsection{Question \theQnum: Can we really achieve global broadband coverage
  directly to smartphones with emerging LEO satellite constellations, or is that a pipe dream?}
\refstepcounter{Qnum} 

Broadband service directly between satellites and cellphones, so-called ``D2C,''
is among the most ambitious proposals for future connectivity.  Apple's surprise
debut in 2022 of a D2C emergency messaging service with its iPhone
14---utilizing the legacy Globalstar LEO constellation---triggered intense
interest and a flood of new investment.  SpaceX (via Starlink) and AST
SpaceMobile (ASTS) aim to provide broadband D2C service to ordinary 5G
smartphones within the next few years.  Amazon (via Kuiper) may be expected to
follow suit. Such proposals, met by extreme skepticism a short time ago
\cite{verizon2020contraAST}, are gaining credibility: even erstwhile skeptics
have now signed partnership agreements for D2C service
\cite{astVerizonPartnershipAnnouncement2024}.  We focus on the D2C question here
since it is already well-understood that fixed or large-form-factor transceivers
can be well-supported by LEO systems such as Starlink.

The current academic literature on non-terrestrial networks (NTN) in the 6G era,
e.g., \cite{giordani2020non, boumard2023technical,guidotti2024role}, highlights
research challenges and standardization efforts, but does not offer accurate
specifics about the expected performance of future D2C services.  Nonetheless,
one can make reasonable projections based on information gleaned from public
filings and initial service test reports.  In our view, the most intriguing D2C
prospect is ASTS, which has been testing its D2C concept since 2019.  Whereas
SpaceX's proposed D2C service is currently based on hardware add-ons to its
existing v2.0-Mini Starlink satellites, and so is constrained to using smaller
antenna arrays, ASTS satellites are purpose-built for D2C and boast enormous
antenna arrays capable of narrow beamforming.  Since ASTS's D2C proposal
represents the most promising scenario for high-rate service from a technical
point of view, let us consider it in a brief case study.

\textbf{D2C Case Study: Will it really work?}  ASTS's Block 2 BlueBird
satellites, whose first launch is scheduled for 2025, are designed with an
extremely large 223-m$^2$ planar area with a phased array on the Earth-facing
side and solar arrays on the other \cite{kookReportOnAst2024}.  Each satellite
is designed to support 2500 adjustable beams, with each beam supporting a
bandwidth of 40 MHz \cite{astArchivedPressRelease2024, kookReportOnAst2024}.  A
few simple calculations are useful to explore the implications of the
satellite's characteristics.  Assuming a wavelength $\lambda = 34$ cm
(corresponding to a downlink frequency of 880 MHz), an antenna area $A = 223$
m$^2$ and efficiency $\eta = 0.8$, the maximum gain of the BlueBird's phased
array is approximately $G_\text{t} = 4 \pi \eta A/\lambda^2 = 43$ dBi, which
implies an approximate beamwidth of $\theta_\text{t} = 0.0255$ rad, or 1.46 deg.
At an altitude of 725 km \cite{boumard2023technical}, the narrowest terrestrial
spot beam that a Block 2 satellite could form would have a diameter of 18.5 km
and an area of 269 km$^2$.  Early testing indicates a spectral efficiency of 3
bps/Hz, which implies that each 40-MHz beam could support a total rate of 120
Mbps on the downlink.  Suppose a beam were directed at a sparse rural area with
population density 30 per km$^2$ and a conservative 50\% smartphone ownership
(the U.S. average in 2023 was 90\%).  This would imply a total of
$269 \times 30 \times 0.5 = 4035$ smartphones in the beam-cell.  Assuming a
conservative 5\% peak concurrency, about 200 of these phones would be active
during peak demand hours.  Even taking the FCC's outdated (2020) definition of
broadband service (25/3 Mbps down/up), fewer than 5 of these 200 active phones
could be supported with a broadband downlink through the 120-Mbps beam.

\textbf{Direct to handset: not broadband, but still revolutionary.}  We conclude
that ASTS's ambitious design could not meet even an outdated definition of
broadband for its D2C service, even in a sparse rural area.  Unfortunately,
further increasing the phased array size, or employing distributed arrays
\cite{tuzi2023distributed}, or multi-satellite MIMO \cite{guidotti2024federated}
offer at best an uncertain path to significantly increasing the D2C area
spectral efficiency.  Yet if expectations were reduced so that 3G-like data
rates---say, 3 Mbps on the downlink---were considered acceptable, then 40 of the
200 peak-active phones in a sparse-rural beam footprint could be served, which
begins to sound viable from both a business and coverage standpoint.  And
assuming high-quality voice calls at 85 kbps with a 5\% concurrency, nearly
30,000 subscribers could be supported within a beam cell, which would satisfy
voice demand even at 90\% smartphone ownership in areas with a population
density up to 125 per km$^2$ (moderately dense rural).

Our overall assessment, therefore, is that D2C services in the 6G era will not
meet any standard definition of broadband but will nonetheless be revolutionary,
providing 3G-like speeds (3-5 Mbps on downlink) to any customer with an
unobstructed view of the sky, and voice or text connectivity even in cars,
trains, or superficially indoors.  D2C will never be data-rate competitive with
terrestrial cellular broadband, whose cost-per-GB to the MNO is about 50x lower
\cite{madden2024ntnEconomics}.  But it is highly cost-effective in terms of
providing true global coverage, since its cost-per-km$^2$ of coverage area is
roughly 1000x lower \cite{madden2024ntnEconomics}.  Thus, D2C can provide
cost-effective infill for customers in rural areas (perhaps via government
subsidy) in the 6G era, as well as a global backup option to customers willing
to pay a premium.

\subsection{Question \theQnum: What role will the 6G cellular standard play in D2C?}
\refstepcounter{Qnum}

A curious fact about D2C is that the companies vying to provide initial D2C
service claim no need of the upgrades to 5G NR that the 3GPP has developed in
Releases 17 and 18 to support NTN.  Instead, the goal is to provide D2C service
to existing pre-Release-17 5G phones and even to LTE phones.  However, the round
trip time (RTT) and carrier frequency offset (CFO) for a direct connection to a
LEO satellite are both well beyond what can be supported by pre-Release-17 3GPP
standards \cite{kodheli2021random,cioni2023physical}.  The trick is to modify
the satellite's base station software---be it in the satellite itself, or at
the ground station (gateway)---to compensate for the additional Doppler and
latency of a D2C connection.  Does this mean that the 3GPP's efforts to
introduce new standards for NTN, including for 6G, are irrelevant?  In the short
term, yes; in the long term, probably not.

One must bear in mind two facts when assessing the need for upgrades to 3GPP
standards to support D2C. First, providers of LEO-based D2C with narrow beams,
such as ASTS, know where their subscribers are to within the beam footprint
radius, or approximately 10 km.  Thus they know the user-equipment (UE) to
satellite latency for all users in a beam to within 20 $\mu$s or so (for
high-elevation-angle connections), and they know the CFO for all UEs in a beam
to better than the uncertainty caused by UE oscillators. 
Second, D2C providers have enormous flexibility in defining proprietary
alterations to base station software.  Thus, for example, none of the challenges
of the random access (RA) procedure identified in \cite{kodheli2021random} are
insurmountable, and none of the fixes posed in \cite{kodheli2021random} and
\cite{cioni2023physical} are absolutely necessary to support high scheduling
efficiency, provided alterations to the gNB software are allowed.  True, the
random access opportunity (RAO) slot identified in the synchronization burst
broadcast by each satellite will not be reachable by a pre-Release-17 UE because
the maximum allowed timing advance is too short to compensate for the UE-to-gNB
latency \cite{cioni2023physical}.  But a modified gNB knows this latency and can
account for it by shifting its acceptance window for the Message 1 preamble
response from the UE.  The only cases that truly pose a challenge are those
involving RTT measured at the UE, such as expiry of the RA response (RAR) window
\cite{kodheli2021random}.  But 5G allows RAR windows of 20 ms or more, which is
forgiving enough to accommodate latencies in LEO-based D2C, even under a
transparent architecture.

In the long term, efforts made in 6G to support NTN will be useful, particularly
for larger beams or higher orbits supporting text and voice services.

We also note that considerable regulatory risk has haunted D2C proposals since
the earliest filings.  The primary question has been whether D2C would only be
allowed on the bandwidth-limited mobile satellite service (MSS) spectrum---the
model Apple has adopted in partnership with Globalstar for its emergency
messaging service---or whether the much more plentiful spectrum owned by MNOs,
until now limited to terrestrial cellular networks, would be opened up for D2C
usage.  The FCC's 2024 Supplemental Coverage from Space (SCS) regulatory
framework resolved this question \cite{fcc2024scsRules}.  It allows MNOs to use
their spectrum for D2C provided that no harmful interference spills into
existing MSS bands or into bands owned by other MNOs.  This has been broadly
interpreted as a green light both for D2C and for the expanded vision of a fully
integrated and seamless terrestrial-and-space cellular communications
architecture such as described in \cite{guidotti2024role}.  Of course,
coordination with 3GPP and other global organizations will be necessary to
establish a global system of rules for D2C and other NTN services.


\subsection{Question \theQnum: What about other NTN paradigms such as HAPS (High
  Altitude Platform Stations) for achieving global coverage?}
\refstepcounter{Qnum}

When Google began development on Project Loon in 2011, it made sense to
investigate whether high-altitude balloons could be a viable platform for 3G
(and later 4G) base stations delivering cellular connectivity to unconnected
areas \cite{loon2021lessonsLearned}.  The space-based alternative was scarcely
considered: 
high launch costs, at about \$16,000 per kg in 2011
\cite{citigroup2022launchCost}, precluded that option in view of the
hundreds---if not thousands---of satellites that would be required.  Other ideas
for providing internet connectivity via high-altitude platform station (HAPS),
such as stratospheric drones, were also the subject of serious development and
investment in the 2010s.  Google acquired Titan Aerospace in 2014, and Facebook
flew its carbon-fiber solar-powered Aquila plane in 2016.

By the time Project Loon was abandoned in 2021, the competitive landscape looked
much different.  SpaceX's reusable Falcon 9 had reduced launch costs to \$2,500
per kg for external customers \cite{citigroup2022launchCost}, and likely below
half this for SpaceX's internal customer, Starlink, whose constellation was
poised to enter commercial service.  Moreover, in 2019 a proof-of-concept
demonstration by AST SpaceMobile showed the feasibility of direct-to-phone links
from LEO \cite{kookReportOnAst2024}.  The reason given for Loon's failure was
lack of commercial viability \cite{loon2021lessonsLearned}, a condition to which
the increasing feasibility and decreasing costs of a space-based alternative no
doubt contributed.  By the time of Loon's demise in 2021, Google and Facebook
had already abandoned their other HAPS programs.  Local and ad hoc NTNs,
including HAPS, will eventually materialize, delivering broadband direct to
phones for special events or regional networks. But there appears to be no
viable path to continental-scale coverage via HAPS or other non-space NTN
paradigms.

\subsection{Question \theQnum: Is O-RAN going to be transformative, or
  will the operators stick with a more vertically integrated
  system?}
\label{sec:ORAN}
\refstepcounter{Qnum} O-RAN makes sense from many perspectives.  By
disaggregating radio access networks and standardizing interfaces between
essential network functions, O-RAN offers operators expanded flexibility in
vendor selection and thus, in principle, lower cost to equip and maintain their
networks \cite{polese2023understanding}.  Moreover, the standardized interfaces
and modules in O-RAN offer insertion points for software virtualization and
incremental adoption of third-party applications that can harness the latest
advances in AI/ML to optimize network configuration and operation
\cite{bonati2021intelligence}.  In fact, a strong case can be made that O-RAN is
key to expanded and agile adoption of AI into the RAN, as it offers a practical
middle way between the slow and highly constrained path to AI adoption via the
3GPP standardization process, and the impractical AI-from-the-ground-up RAN
model that eschews standardization (see Sec. \ref{sec:MLPHY1}).  With such
compelling arguments to recommend it, the O-RAN architecture is commonly viewed
as a defining feature of future 6G networks \cite{lin2023embracing5g6g} and one
of the keys to expanding coverage. But it should be noted that some major
players, such as Huawei, are not a part of existing O-RAN development or lobbying
groups, and are instead deploying alternative solutions. We expect tight
coordination between 3GPP and O-RAN in standardizing the 6G architecture.

We believe RAN disaggregation will become the dominant paradigm going forward
whether or not it gives rise to significant vendor diversification.  The
functional splitting initiated in 5G between distributed units (DUs) and
centralized units (CUs) of the RAN makes sense even for single-vendor systems,
because high-performance yet inexpensive off-the-shelf (OTS) servers can replace
all Layer-2 functions.  Virtualization and modularization will, in turn,
significantly erode the ``G staircase'' as discussed in Sec. \ref{sec:endofGs}:
the software modules in the DU and CU will be subject to continuous development,
shadow testing, and deployment schedules with fine-grained versioning, just as
is common in the software industry. Again, these benefits will no doubt be
realized even if vendors maintain monolithic ownership of the RAN stack.

\textbf{``Open'' comes with costs: complexity, performance, and security.}
O-RAN faces significant headwinds that arise precisely because of its open,
modular, and clearly defined nature.  The first challenge is
complexity. Operators of networks based on O-RAN will either have to assume the
complex burden of system integration themselves, or they will need to hire
system integrators whom they will hold accountable to provide smooth network
operation despite a heterogeneous mix of vendors and equipment.  The stringent
demands of high performance and extreme reliability will induce operators or
integrators to settle on a small subset of well-tested multi-vendor
combinations, partially vitiating the open competition that originally motivated
O-RAN as a concept.

The second challenge is performance.  Multi-vendor systems will likely struggle
to achieve the same performance as sole-vendor disaggregated RANs because the
latter can pass additional proprietary messages from one module to the next,
enabling cross-module optimization, whereas multi-vendor systems will be tied
exclusively (or nearly so) to O-RAN-defined interfaces.  Moreover, systems
running on a cascade of proprietary ASICs will tend to be more energy
efficient---pehaps significantly so---than those based on virtualization across
OTS servers.  Thus, costs saved by implementing processing on OTS servers could
be lost in the long term to greater energy expenditure.

O-RAN's third challenge is security.  Open interfaces and modularity are
well-known to expand a system's attack surface, and O-RAN is no exception
\cite{polese2023understanding, xing2024fronthaulSecurity}.  In an industry
already fraught with suspicion over eavesdropping and data
manipulation---including outright bans on several major equipment vendors in
many Western countries---national authorities are unlikely to welcome a
flourishing of third-party apps running on open multi-vendor platforms drawing
from global supply chains.

\textbf{O-RAN will succeed, but conditionally.}  Our assessment is that RAN
disaggregation is inevitable---indeed, already well underway---but that only a
small set of trusted vendors and system integrators will ultimately succeed in
the 6G O-RAN economy.  Third-party rApps and xApps will indeed run on RAN
intelligent controllers (RICs), but like apps on Apple or Android platforms,
they will be tightly contained for security reasons, depriving them of access to
data streams that could enhance their performance, whereas proprietary apps from
the vendor or system integrator will have freer access.  Thus, disaggregation
will ultimately usher in a new age of AI-infused modular RANs, some running on
OTS equipment, but vendor (or system integrator) lock-in will remain a feature
of 6G because performance, energy efficiency, and trust will remain imperatives
for operators and regulators.

\section{Acknowledgments}

The authors would like to thank Alperen Duru for his help creating the figures in this article.  

\bibliographystyle{ieeetr}
\bibliography{Andrews,pangea,Tingfang}
\vspace{0.5in}

\noindent \textbf{Jeffrey Andrews (F'13)} is the Truchard Family Endowed Chair
in Engineering at the University of Texas at Austin where he is Director of the
6G@UT research center, which promotes close collaboration with over a dozen leading companies in the 6G ecosystem.  Dr. Andrews is an ISI Highly Cited Researcher and has been co-recipient of 16 best paper awards in the field of wireless communications, including the 2016 IEEE Communications Society \&
Information Theory Society Joint Paper Award and six other major IEEE journal paper awards.  He has received the the 2015 Terman Award, the 2021 Gordon Lepley Memorial Teaching Award, and the 2019 IEEE Kiyo Tomiyasu technical field award.  He was the Founding Chair of the Steering Committee for the IEEE Journal on Selected Areas in Information Theory and is a past Editor-in-Chief of the IEEE Transactions on Wireless Communications.  He received the B.S. in Engineering with High Distinction from Harvey
Mudd College, and the M.S. and Ph.D. in Electrical Engineering from Stanford
University.  \vspace{2mm}

\noindent \textbf{Todd Humphreys (SM)} holds the Ashley H. Priddy Centennial
Professorship in Engineering in the department of Aerospace Engineering and
Engineering Mechanics at the University of Texas at Austin.  He is Director of
the Wireless Networking and Communications Group (WNCG) and of the UT
Radionavigation Laboratory.  His publications are among the most cited in the
field of positioning, navigation, and timing.  His awards include The University
of Texas Regents' Outstanding Teaching Award (2012), the National Science
Foundation CAREER Award (2015), the Institute of Navigation Thurlow Award
(2015), the Qualcomm Innovation Fellowship (2017), the IEEE Walter Fried Best
Paper Award (2012, 2020, 2023), the Presidential Early Career Award for
Scientists and Engineers (PECASE, 2019), and the Institute of Navigation Kepler
Award (2023). He is a Fellow of the Institute of Navigation and of the Royal
Institute of Navigation.  He earned his B.S. and M.S. in Electrical and Computer
Engineering from Utah State University, and his Ph.D. in Aerospace Engineering
from Cornell University.  \vspace{2mm}

\noindent \textbf{Tingfang Ji} is Vice President of Engineering at Qualcomm Technologies, Inc.,
where he leads global 6G research. He joined Qualcomm Technologies in 2003, and from 2003 to 2014, he made instrumental technical contributions toward the development of LTE and LTE-Advanced technology and served as an official of 3GPP radio group. Since
2014 he has been responsible for the flagship Qualcomm Technologies 5G/6G research project, driving Qualcomm Technologies' 5G NR air interface design and standardization efforts, sub6 GHz multi-vendor pre-commercial IODT/trials, and 6G research on communication theory, positioning, sensing, full duplex, and cross-node wireless machine learning. Tingfang is Chair of the technology working group of NextG Alliance, which promotes North American 6G technologies. Before joining Qualcomm, Tingfang was a member of the technical staff at Bell Labs. As an inventor, he has more than 600 granted US patents.  Tingfang received his Ph.D. in E.E. from the University of Michigan, Ann Arbor, in 2001 and a B.Sc. from Tsinghua University,
Beijing.\vspace{2mm}

\end{document}